\documentclass[aps,prb,twocolumn,showpacs,amsmath,amssymb,superscriptaddress
]{revtex4-1} 
\usepackage[english]{babel}
\usepackage{amsmath,amssymb,amsfonts} 	
\usepackage{graphicx}
\usepackage[utf8]{inputenc}
\usepackage{bm}
\usepackage{xcolor}
\usepackage[normalem]{ulem}
\usepackage{SIunits}
\usepackage[colorlinks=true,linkcolor=blue,citecolor=blue,urlcolor=blue]{hyperref}
\usepackage{url}
\usepackage{textcomp}
\usepackage{lineno,hyperref}
\usepackage{float}
\usepackage{chemfig}
\usepackage{rotating}
\usepackage{subcaption}
\usepackage{color} 
\definecolor{red}{rgb}{1.,0.,0.}
\usepackage{soul}
\usepackage{amsmath}

\usepackage{graphicx,wrapfig,lipsum}
\usepackage{caption}
\DeclareUnicodeCharacter{00A0}{~}
\makeatletter
\newcommand*{\rom}[1]{\expandafter\@slowromancap\romannumeral #1@}
\makeatother

\usepackage[inline]{asymptote}

\begin{asydef}
// Global Asymptote definitions
real linkLen=1, linkWidth=2pt;
real rl=2+linkLen;              // distance between beads
guide link=(1,0)--(1+linkLen,0);   // a link
pen beadColor=orange;
pen linkColor=beadColor;
void bead(transform t){
  draw(t*link,linkColor+linkWidth);
  radialshade(t*unitcircle,
    beadColor,shift(t)*(-0.4,0.3),0.01
   ,black,shift(t)*(-0.4,0.3),1.5);
}
pair operator>(pair pos=(0,0), real phi){
  transform t=shift(pos)*rotate(phi);
  bead(t); // draw a bead with a link
  pos+=rl*(Cos(phi),Sin(phi)); // Sin, Cos - in degrees, sin, cos - in radians
  return pos;
};
pair pos;
\end{asydef}
\usepackage{tikz}
\usepackage[labelfont=bf,
justification=raggedright]{caption}
\newcommand\THEOSMARVEL{Theory and Simulation of Materials (THEOS) and National Centre for Computational Design and Discovery of Novel Materials (MARVEL), {\'E}cole Polytechnique F{\'e}d{\'e}rale de Lausanne, 1015 Lausanne, Switzerland}
\newcommand\PSI{Laboratory for Materials Simulations, Paul Scherrer Institut, 5232 Villigen PSI, Switzerland}

\newcommand\Montpellier{Laboratoire Charles Coulomb (L2C), Université de Montpellier, CNRS, Montpellier, France}

\graphicspath{{./images_paper/}}
\begin{document}

\title{Infrared-active phonons in one-dimensional materials and their spectroscopic signatures}
\author{Norma Rivano}
\affiliation{\THEOSMARVEL}

\author{Nicola Marzari}
\affiliation{\THEOSMARVEL}
\affiliation{\PSI}

\author{Thibault Sohier}
\affiliation{\Montpellier}

\begin{abstract}
Dimensionality provides a clear fingerprint on the dispersion of infrared-active, polar-optical phonons. 
For these phonons, the local dipoles parametrized by the Born effective charges drive the LO-TO splitting of bulk materials; this splitting actually breaks down in two-dimensional materials. Here, we extend the existing theory to the one-dimensional (1D) case.
Combining an analytical model with the implementation of density-functional perturbation theory in 1D boundary conditions, we show that the dielectric splitting in the dispersion relations collapses logarithmically at the zone center. The dielectric properties and the radius of the 1D materials are linked by the present work to these red shifts, opening novel IR and Raman characterization avenues.
\end{abstract}

\maketitle

\noindent

Phonons and their interactions with electrons and photons are key ingredients in determining the thermodynamic, transport, and optical properties of materials \cite{ziman2001electrons,ashcroft1976solid}. 
Notably, long-wavelength optical phonons can give rise to electric fields which strongly affect not only their dispersion relations \cite{sohier2017breakdown,PhysRevX.11.041027, mele2002electric, sanchez2002vibrational, michel2009theory, michel2009theory, de2020experimental}, but also the physics of Fr\"{o}hlich electron-phonon interactions \cite{sohier2016two,PhysRevB.92.054307,verdi2015frohlich} and phonon polaritronics \cite{ rivera2019phonon}.
In semiconductors and insulators, when atoms in the lattice have non-vanishing Born effective charges (BECs), optical phonons can generate a polarization density and couple with electric fields. Those modes are then termed polar and are infrared (IR) active. In addition, for longitudinal atomic displacement patterns, and in particular for purely longitudinal optical (LO) modes, a long-range (LR)  electric field is generated that becomes macroscopic in the long-wavelength limit \cite{born1954k,PhysRev.59.673, pick1970microscopic}.
Creating an additional electric energy density in the material is more costly, and thus the frequency of the LO mode is blue-shifted.
While the strength of this effect depends on the dielectric properties of the material (BECs and the high-frequency limit of the dielectric tensor $\epsilon_\infty$), its dependency on phonon momenta and size is ruled solely by dimensionality, and we argue here that this unique fingerprint on the dispersion relations can be exploited for spectroscopic characterization or in opto-electronic devices.
In 3D, the shift in the energy of the LO mode is constant around the Brillouin zone (BZ) center as a function of the norm of the momentum. At variance, in 2D it has been shown to depend linearly on momentum and to vanish at $\Gamma$ exactly\cite{sohier2017breakdown,PhysRevX.11.041027, zardo2009raman}.
This breakdown can be expected in 1D systems as well\cite{zhang2008long, piscanec2007optical,  adu2006raman, mele2002electric, nemanich1981light}; nonetheless, its actual behavior remains an open question.

The dielectric contribution to the dispersion of the LO mode is often described in terms of a deviation from the transverse optical (TO) mode: LO-TO splitting.
This is because in many materials (e.g., with cubic/tetragonal symmetries or planar hexagonal) LO and TO modes would be degenerate in the absence of dielectric effects \cite{cochran1962dielectric, pick1970microscopic, baroni2001phonons,born1954k, baroni2001phonons, giannozzi1991ab, gonze1997dynamical}.
However, the lifting of these degeneracies ultimately depends on the symmetries and the dimensionality of the crystal.
In 3D, with 3 equivalent directions at most, optical modes are up to triply degenerate at the zone center (based on group theory considerations \cite{tinkham2003group}).
In 2D, these modes are up to doubly degenerate, while a  splitting with respect to the out-of-plane optical (ZO) modes always persists since  in- and out-of-plane displacements are  nonequivalent. 
In 1D, the longitudinal direction is clearly different from the other two, possibly degenerate with each other. Thus, there is no degeneracy to recover, even if the polar energy shift vanishes.
Accordingly, we will speak of dielectric or polar shift rather than LO-TO splitting.

In this work, we investigate IR-active phonons, developing for 1D the analysis made \cite{giannozzi1991ab, baroni2001phonons, gonze1997dynamical} for 3D \cite{cochran1962dielectric} and 2D materials \cite{sohier2017breakdown,PhysRevX.11.041027}.
We implement density-functional perturbation theory (DFPT) \cite{baroni2001phonons} with  1D open-boundary conditions (OBCs)\cite{SI,PRB} to capture the response of isolated 1D systems; we then derive an analytical model describing the interplay between the phonon-induced polarization and electronic screening. Notably, such model enables to interpret Raman and IR spectra of 1D systems, thus greatly aiding material characterization at the nanoscale.
We investigate the dispersion characteristics of prototypical systems such as BN atomic chains, BN nanotubes, and GaAs nanowires.
Remarkably, we show the collapse of the dielectric shift at $\Gamma$ and we derive its logarithmic asymptotic behavior as a function of phonon wavevector $q_z$ and radius $t$ of the material.
This analysis allows to interpret and predict vibrational properties as a function of dimensionality, paving the way to similar developments for optical and transport properties.
As an example, we show how the radius of a 1D system could be extracted from IR or Raman experiments.

We shall frame the discussion by introducing an electrostatic model for the electric field generated by LO phonons, and discuss its consequences on their dispersion relations.
Here, the 1D system is described as a charge distribution periodic along the $z$-axis and homogeneous in the radial direction within an effective radius $t$, with vacuum outside.
Within a dipolar approximation, the atomic displacement pattern $\mathbf{u}_{\nu}^a$ associated with a phonon $\nu$ of momentum $\mathbf{q} = q_z \mathbf{\hat{z}}$ induces a polarization density

\begin{equation}
  \mathbf{P}(q_z)=\frac{e^2}{L}\sum_{a} {Z_a} \cdot \mathbf{u}^a_{\nu} (q_z)  \,,
\end{equation}

where $e$ is the unit charge, $L $ is the unit-cell length, and $Z_a$ is the tensor of BECs associated to each atom $a$ within the unit cell.
The corresponding charge density $\mathbf{q}\cdot\mathbf{P}(q_z)$ is the source of the electric field, termed Fr\"ohlich due to the related electron-phonon interaction, and vanishes as soon as the phonon propagation (along momentum $\mathbf{q}= q_z \mathbf{\hat{z}}$) and the polarization (along displacements $\mathbf{u}^a_{\nu}$) are orthogonal.
Thus, within this model, only phonons labeled as LO in the long-wavelength limit generate a sizable electric field and experience the dielectric shift. In the following, we focus on strictly in-chain atomic displacements $\mathbf{u}^a_{\nu} \to \mathbf{u}^a_{LO}$ and assume BECs and the macroscopic dielectric tensor to be diagonal. 

By solving the associated Poisson equation, we derive the new electrical forces and the resulting change to the phonon frequencies. The full derivation is reported in the SI \cite{SI}. 
After some manipulation, the general expression for a  material in $n$-dimensions can be recast in the form
\begin{equation}\begin{split}
    \omega_{\mathrm{LO}}
    =\sqrt{\omega^2_{0}+\Delta\omega_{\mathrm{max}}^2 \Bigl[1- \Delta_{\mathrm{nD}}(\mathbf{q},t)\Bigr]}\,,
    \label{omega_1d}
    \end{split}
\end{equation}
where $\omega_{0}$ is the reference value for the LO branch in the absence of any additional contribution from polarity.
To ease the comparison between dimensionalities, we have highlighted  two main contributions: $\Delta \omega_{\mathrm{max}}^2$ and  $\Delta_\mathrm{nD}$.
The prefactor $\Delta \omega_{\mathrm{max}}^2= \frac{ 4 \pi e^2}{\epsilon^{m}_{z} \Omega} (\sum_a\frac{ Z_a\cdot\mathbf e^a_{\mathrm{LO}} }{ \sqrt{M_a }})^2$ corresponds to the maximum value of the shift set by the dielectric properties of the nD crystal. Here, $\Omega$ is the volume of material in a cell, i.e., the volume of the unit cell in 3D, or the cell area times the thickness in 2D, or the cell length times the section in 1D. $\epsilon^m_{z}$ is the dielectric tensor component for the propagation direction $\mathbf{\hat{z}}$ and $\mathbf{e}^a_{LO}$ is the LO eigenvector for an atom $a$ scaled by its  mass $M_a$. This prefactor is then modulated by $\Delta_\mathrm{nD}$, whose expression depends on dimensionality: it is zero in 3D, while dependent on phonon momenta (in-plane or in-chain) and size  $t$ (thickness or radius) in 2D and 1D.
The 1D fingerprint derived from the model presented here (assuming the diagonal dielectric tensor to be isotropic,  i.e., $\epsilon^m\rightarrow\epsilon_{\mathrm{1D}}\mathbb{I})$ reads

\begin{widetext}
\begin{equation}
\Delta_{\mathrm{1D}}(q_z,t)=2I_1 (|q_z| t) K_1 (|q_z| t) \Bigl(1 - \frac{2 \epsilon_\mathrm{1D} \sqrt\pi q_z t I_1(|q_z| t) K_0(|q_z| t) -G^{2 2}_{2 4}(|q_z|^2 t^2)}{2\sqrt\pi q_z t(\epsilon_\mathrm{1D} I_1(|q_z| t)K_0(|q_z| t)+I_0(|q_z| t)K_1(|q_z| t))}\Bigr)\,,
\label{Delta}
\end{equation}
\end{widetext}

where $I_n(x)$, $K_n(x)$  are  the $n^{th}$-order modified cylindrical Bessel functions, and $G_{p q}^{m n}\left(
\begin{array}{c}
a_1,...,a_p\\
b_1,...,b_q
\end{array}\middle\vert x\right)$
is the Meijer G-function.
The limit behavior of Eq. \ref{Delta} in the vicinity of $\Gamma$ is  $\Delta_{\mathrm{1D}}(q_z,t) =1-\frac{q_z^2t^2}{2}(C(\epsilon_\mathrm{1D})- \epsilon_\mathrm{1D}\log(q_zt))$, where $C(\epsilon_\mathrm{1D})$ is a constant (independent of $q_zt$) and is reported in the SI\cite{SI}. 
Note that the equivalent in 2D would be\cite{sohier2017breakdown}
$
\Delta_{\mathrm{2D}}(\mathbf{q_p},t)=1- \frac{\epsilon_{\mathrm{2D}}t|{\mathbf{q}_p}|}{2+\epsilon_{\mathrm{2D}}t{|\mathbf{q_p}|}}\,,
$
leading in that case to a linear collapse of the dielectric shift in terms of in-plane phonon momentum $\mathbf{q}_p$.

Eqs. \ref{omega_1d} and \ref{Delta} are the central analytical result of this Letter. 
In particular, $\Delta_{\mathrm{1D}}$ dictates the transition from a momentum-dependent (1D-like) to a momentum-independent (3D-like) shift, similar to what is observed in 2D \cite{sohier2017breakdown}.
In the $q_z t \rightarrow 0$ limit, $\Delta_{\mathrm{1D}} \to 1$ and the shift breaks down at $\Gamma$ with an overbending at small but finite $q_z$ \cite{zhang2008long, piscanec2007optical, adu2006raman, mele2002electric, nemanich1981light}.
Instead, if considering the opposite $q_zt \to \infty$ limit, the modulation typical of low-dimensionality vanishes  and one is left with the well-known constant 3D shift. 
The reason behind this transition is intuitive.
For small perturbing momenta (real-space LR interactions), the electric-field lines associated with the polarization density spread far away in the surrounding medium, leading to vanishing dipolar interactions and shift: the material perceives itself as an infinitely thin 1D system surrounded by vacuum.
As the momentum increases (short range in real space), these lines get more and more confined within the material and the dipole-dipole interactions eventually resemble those of a bulk material, insensitive to the boundaries.

We now combine our analytical findings with first-principles calculations.
In this endeavour, DFPT \cite{baroni2001phonons} represents a valuable ally, although some modifications are required when dealing with low dimensionality \cite{sohier2015density,sohier2016two,sohier2017breakdown,rozzi2006exact,PhysRevX.11.041027,castro2009exact}. The major one stems from periodic-boundary conditions (PBCs) leading to spurious LR Coulomb interactions between periodic images. When the electronic charge density is perturbed at momentum $q$, the reach of these interactions scales as $\lambda= 2\pi/q$ in the out-of-chain (or plane) directions. It follows that, for long-wavelength perturbations, these cross-talks persist even for very large distances, turning the response of the isolated 1D system into the one of a fictitious 3D system of periodic replicas. 
For a systematic and physical solution to this issue, we have implemented a 1D Coulomb cutoff technique, based on the version proposed in Ref.  \cite{rozzi2006exact}, in the relevant packages (PWScf and PHonon) of the Quantum ESPRESSO distribution \cite{RevModPhys.73.515, giannozzi2009quantum,giannozzi2017advanced}.
This implementation is summarized in the SI \cite{SI} while fully detailed in an upcoming publication \cite{PRB}, and restores the physical OBCs for the computation of total energies, forces, stress tensors,  phonons, and electron-phonon interactions. 

\begin{figure*}[t!]
\centering
\includegraphics[scale=0.5]{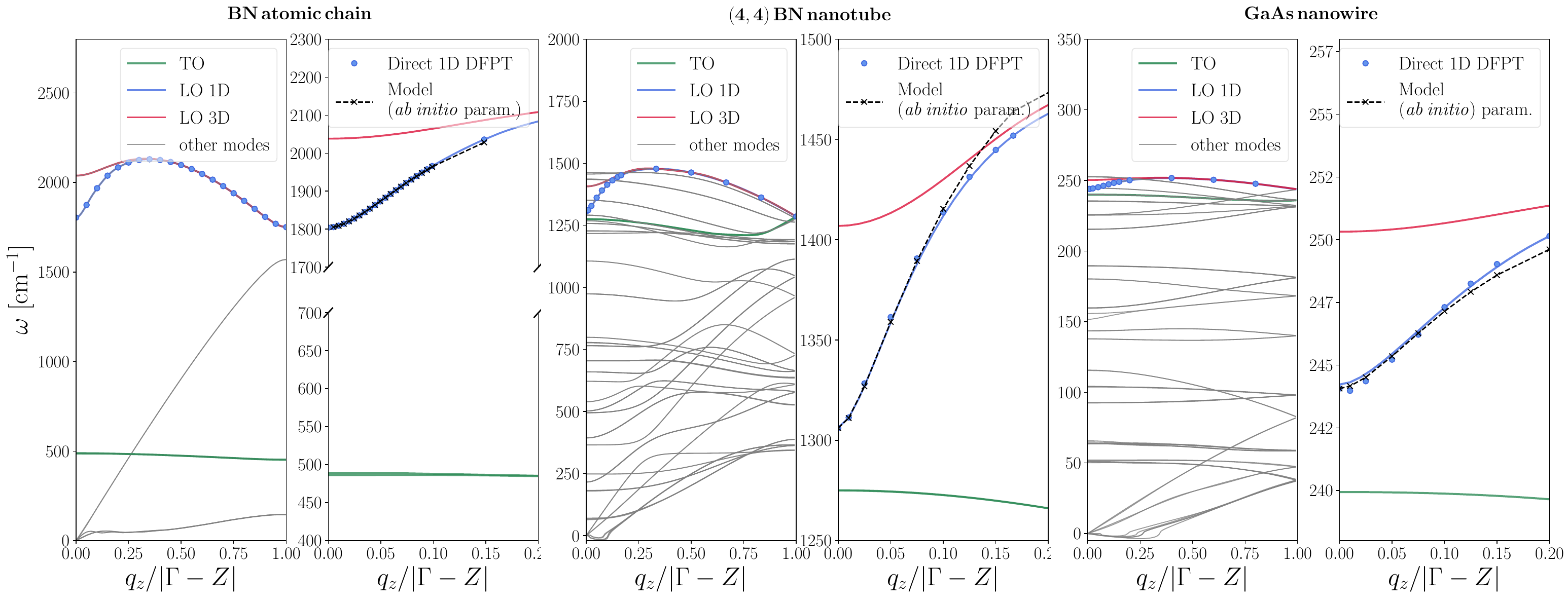}
\caption{Phonon dispersion of BN atomic-chain, (4,4) nanotube, and wurtzite GaAs nanowire of 24 atoms.
For each material, left panel compares 3D-PBC and 1D-OBC DFPT calculations, explicit for 1D (symbols) and interpolated  for both (lines). Right panel shows the agreement of our model with 1D DFPT for the LO branch in the long-wavelength limit.}
\label{BN_atomic-chain}
\end{figure*}

In the following, we focus on  BN atomic chains, BN armchair nanotubes, and GaAs nanowires. 
The scope is threefold: to show the relevance of the OBCs for linear response in 1D, to validate Eq. \ref{Delta}, and to discuss the underlying physics and the transition between dimensionalities. 
The details of the calculations and model parametrization (also independently obtained via 1D DFPT) are given in the SI \cite{SI}.
For each material we compute the phonon dispersions and we plot it in full in the left panel, zooming around the LO-TO splitting in the right panel (see Fig. \ref{BN_atomic-chain}).
We focus on the modes similar to those of the bulk 3D or 2D parents, identifying the purest longitudinal mode (hLO), highest in energy and associated to the largest polarization density, as well as the corresponding tangential or bulk-like transverse modes.
For clarity, colors are meant to highlight these phonon branches, while the others are left in gray in the background.

Fig. \ref{BN_atomic-chain} shows the effect of the spurious interactions between periodic images.
In 3D DFPT (i.e., 3D PBCs), we always recover a rather flat LO branch (red) with a finite dielectric shift at $\Gamma$, which possibly adds up to the splitting with respect to TO phonons (green) due to crystal symmetries and dimensionality.  This is the 3D response of an array of interacting 1D materials. 
On the contrary, with 1D DFPT (i.e., 1D OBCs), the amount of energy built by LO phonons (blue) is shown to vanish  at the zone center and the branch exhibits a logarithmic overbending in the long-wavelength limit: this is the response of the isolated 1D material given by Eq. \ref{Delta}.
The correction introduced by the Coulomb cutoff is shown to be significant for small $q_z$, where low-dimensionality comes into play. 
The BZ range over which the discrepancy between 3D and 1D DFPT extends is determined by the amount of vacuum in the simulation cell. 
In PBCs, the larger the vacuum, the smaller the region affected by the stray fields, and the softer the LO branch; this latter asymptotically converges to the 1D limit. 
The true physical behavior is fully recovered only in the presence of the cutoff, since for momenta smaller than the inverse of the distance between periodic images there will always be the response of a 3D periodic system, i.e., a non vanishing polar shift.

In the right panels of Fig. \ref{BN_atomic-chain} we present the comparison between 1D DFPT and the analytical model we have derived.
For all materials, an excellent agreement is found at the very least within the first $10\%$-$20\%$ of the BZ, that is the long-wavelength limit targeted by the model.
The strength of the effect, being the range of frequencies over which the overbending extends, is material dependent and ruled by the square of the screened effective charges, i.e., $\frac{Z_a^2}{\epsilon_\mathrm{1D}}$ (see Eq. \ref{omega_1d}).
By comparing the materials in Fig. \ref{BN_atomic-chain}, the polar shift is obviously most pronounced in BN: around $200$ $\mathrm{cm^{-1}}$ in the case of nanotubes, consistently with 2D and 3D hexagonal BN \cite{sohier2017breakdown,rokuta1997phonon,geick1966normal}, and around $400$ $\mathrm{cm^{-1}}$ for the chain.
In the BN chain, the larger increase is due to the crystal structure differing from the hexagonal one common to the other allotropes.
The effect is more subtle in GaAs, of about 10 $\mathrm{cm^{-1}}$, because of the significantly smaller BECs\footnote{As a consequence of the pseudopotentials used in this work, GaAs frequencies are strongly underestimated. The actual range of the polar-effect should be approximately $20 \, \mathrm{cm^{-1}} $, as confirmed by our tests with other pseudopotentials \cite{SI}. However, here the focus is mainly on the qualitative trend of the polar effect more than the exact frequencies. Note that the suitable pseudopotentials are used for comparison with experiments in the following section. }.
 
In low-dimensional materials, as a consequence of the vanishing polar shift, the remaining LO-TO splitting at the zone center is  purely ‘mechanical’, i.e., due to structurally different atomic displacements as a consequence of symmetry and dimensionality. 
Among the selected materials, the chain represents the ultimate 1D system and exhibits the largest mechanical splitting (i.e., larger asymmetry between displacement directions). 
Instead, nanotubes and nanowires sit in between 1D and 2D and 3D, respectively, as a function of their diameter. 
The mechanical splitting at $\Gamma$ is expected to decrease as $ t \rightarrow \infty $, converging to the 2D or 3D case. Similarly, the polar shift asymptotically converges to its higher-dimensional limit. Here,  \lq asymptotically\rq \, is key, since the 1D nature of the material will always suppress the polarity-induced electric field at small enough momenta. 
Thus, the effect of the radius increase is visible in the long-wavelength regime: 
the range of momenta over which the shift vanishes shrinks and the discontinuity at $\Gamma$ due to  direction-dependent BECs is transferred from the prefactor of the logarithmic overbending (1D) to the slope\cite{sohier2017breakdown} (2D) or the value\cite{cochran1962dielectric, pick1970microscopic, baroni2001phonons,born1954k, baroni2001phonons, giannozzi1991ab, gonze1997dynamical} (3D) of the polar shift.

\begin{figure}[t!]
    \centering
      \includegraphics[width=0.5\textwidth]{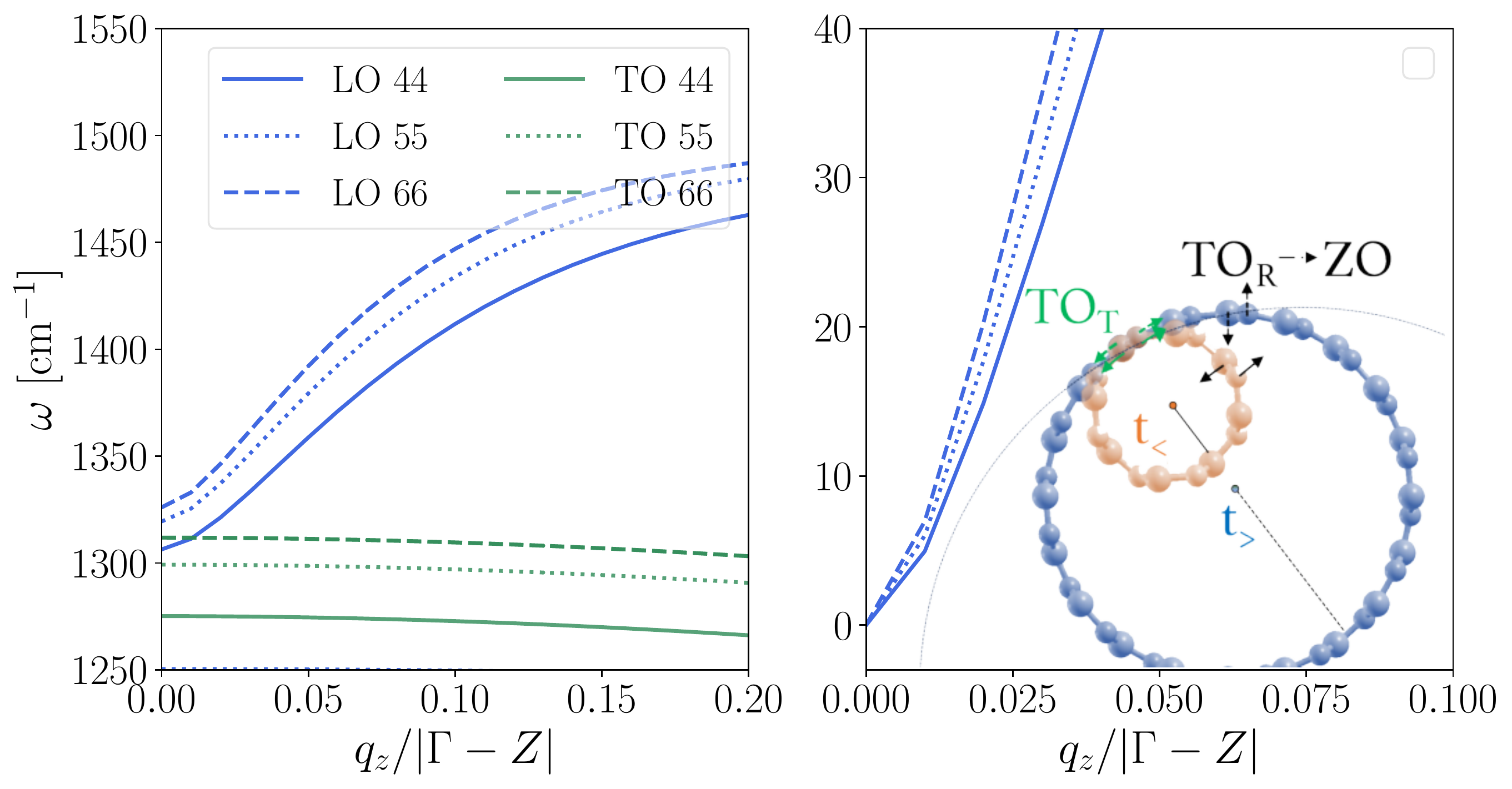}
    \caption{Size effects (mechanical and polar) on LO and TO modes for BN (6,6), (5,5) and (4,4) armchair nanotubes from 1D-DFPT. Right panel compares both modes.
    Left panel focuses on the polar shift for the LO branch by setting as common offset $\omega_0=0$. The sketch represents the mechanical evolution of tangential $\mathrm{TO_T}$ vs radial $\mathrm{TO_R}$ optical modes.}
    \label{size_bn-nanotubes}
\end{figure}

Focusing on nanotubes, Fig. \ref{size_bn-nanotubes} compares the first-principles results for (4,4), (5,5) and (6,6) BN tubes. In the right panel, decreasing the curvature is shown to stiffen the logarithmic LO behavior and approach the linear signature of 2D materials\cite{sohier2017breakdown}.
The left panel shows instead the absolute values of the optical frequencies for each tube and focuses on the mechanical size effects.
There, one can observe two trends.
On one hand, it is known throughout the literature that increasing the radius mechanically blue-shifts both TO and LO phonons at zone center \cite{mahan2003optical, PhysRevB.74.033407, campbell1986effects, richter1981one,xiong2006raman,fauchet1988raman, zhang2008long, zardo2009raman, PhysRevB.68.045425, sanchez2002vibrational, popov2003lattice, adu2006raman,  piscanec2007optical}.
On the other hand, the same radius increase progressively reduces the mechanical splitting,
converging towards a finite or null gap depending on the symmetries of the 2D parent (as sketched in Fig. \ref{size_bn-nanotubes} for the tangential modes).
Note that the mechanical shifting is much stronger for the TO mode, which can be understood by considering that the atoms are displaced in the non-periodic direction. Furthermore, since the mechanical contribution to the LO shift actually opens the LO-TO gap (pushing LO up), the closing of the gap can be attributed to the stronger increase of the TO frequency.
A similar analysis holds for nanowires as well, but in this case the transition is from 1D to 3D. 
The mechanical effects are shared with nanotubes, depending only on crystal symmetries. As regards the dielectric shift, the range of momenta over which the splitting goes from zero to its constant 3D value progressively shrinks around $\Gamma$, the dispersion becoming progressively steeper and only asymptotically approaching the well known discontinuity in the bulk.

Once the relevant  parameters  in Eqs. \ref{omega_1d}, \ref{Delta} are known, empirically or from first-principles, the model unveils a one-to-one relationship between first-order LO Raman/IR lines (i.e., frequency) and radius $t$. 
This makes the model a very valuable tool for spectroscopic characterization of 1D systems. 
The phonon probed in experiments is with small but finite momentum $q_z$, related to the laser wavelength $\lambda$ by momentum conservation law (i.e., $q_z=2\pi/\lambda$).
Thus, the larger the $\lambda$, the closer to $\Gamma$ is the $q_z$ probed.
The evolution with size of the hLO branch is given by Eq. \ref{omega_1d}, where size effects are conveyed by the change in radius explicitly appearing in the $\Delta_\mathrm{1D}(q_z,t, \epsilon_\mathrm{1D})$.
In passing, note that the dielectric properties and the absolute positions of LO/TO modes may actually change with size due to mechanical reasons, as well as the q-dependency of the eigenvectors \cite{SI}. 
We emphasize that the roles of $t$ and $q_z$ in $\Delta_{1D}$ are symmetric (i.e., $\Delta_{1D}$ depends on the product $q_zt$). 
Thus, the behavior of the frequency versus size $\omega_\mathrm{LO}(t)$ at fixed phonon momenta is the same as the phonon dispersion $\omega_\mathrm{LO}(q_z)$ at fixed radii. 
The polar shift increases logarithmically at small $t$, then approaches a maximum set by the bulk splitting. The increase is sharper for smaller wavelengths, meaning that a larger $\lambda$ would instead ease size resolution \cite{SI}.

Raman/IR experiments on single, isolated and semiconducting wires/tubes  are mostly missing, and closest to the conditions discussed here is the work of Ref. \cite{kim2017bistability}, where the authors  propose a new strategy to grow  ultrathin GaAs nanoneedles with atomically sharp tips ($t\approx$ 10 nm, mostly wurtizte) on top of thicker bases ($t\approx$ 100 nm, zincblend). The nanowires are arranged in regular arrays with a spacing of 1000 $\mathrm{nm}$. 
This distance, compared with the laser spot, is large enough to avoid significant inter-wire cross-talks hindering the response of the single wire. 
Then, size effects are investigated with room-temperature Raman spectroscopy by pointing a semiconducting laser probe at 785 $\mathrm{nm}$ on either the tip or the base. 
The observed spectra are composed of two peaks each. For the base, the TO and LO modes are found at 268 $\mathrm{cm^{-1}}$ and 285 $\mathrm{cm^{-1}}$ (LO-TO splitting $\approx$ 17 $\mathrm{cm^{-1}}$), respectively. The tip spectrum, instead, systematically exhibits, besides broadening,  a stable TO mode and a down shift of about 3 $\mathrm{cm^{-1}}$ for the LO mode.
The observed change in the LO-TO separation, switching from the base to the tip, appears to be almost entirely due to the change in LO position.
A mechanical red shift  is expected to affect equally the two modes or mostly the TO, which is instead stable.
This points to the dielectric nature of the phenomenon, i.e., the vanishing polar shift and its change with size.

\begin{figure}[t!]
    \centering
    \includegraphics[width=0.5\textwidth]{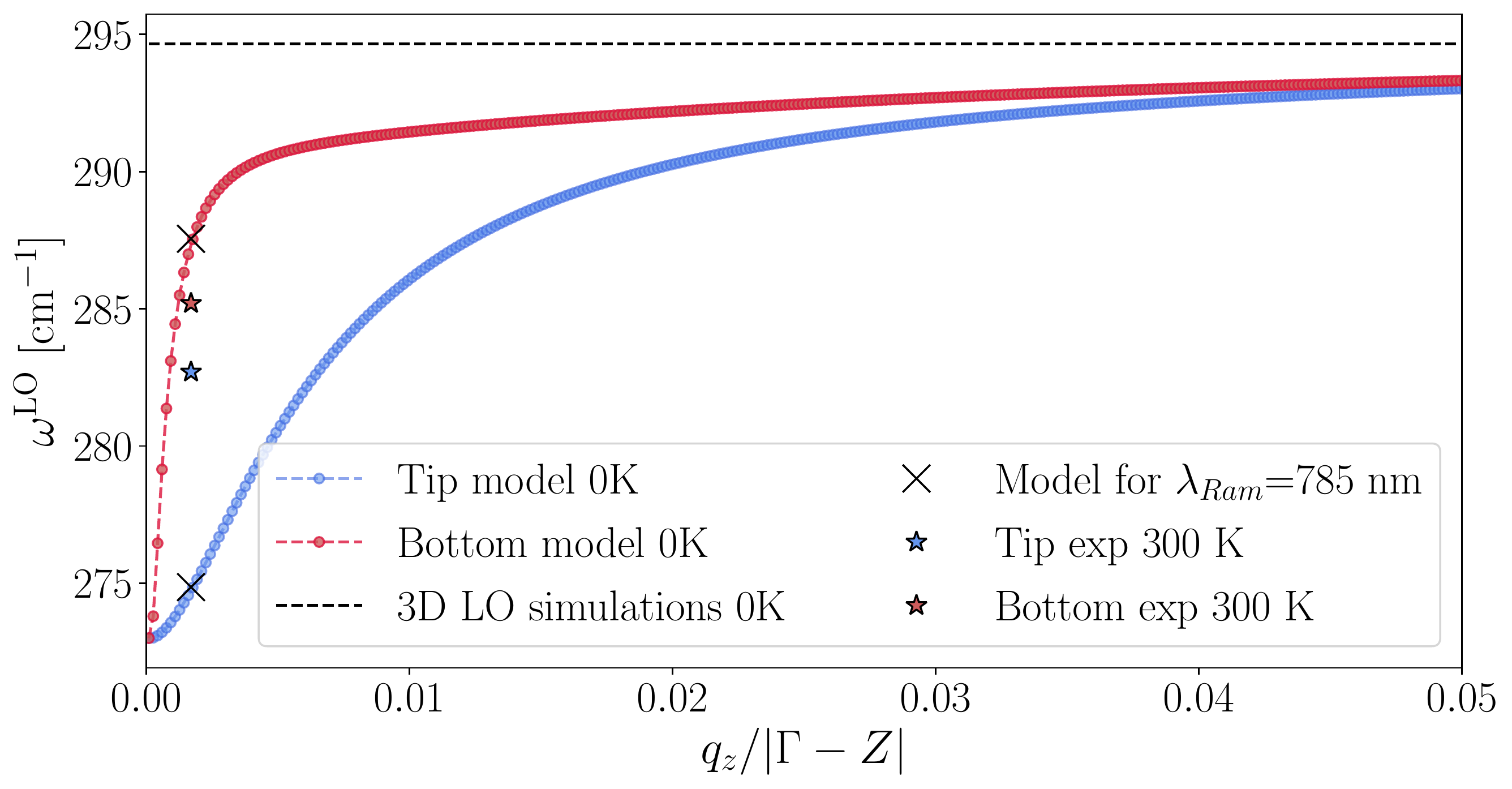}
    \caption{Evolution of the LO branch in the long-wavelength limit for the two GaAs nanowires from Ref. \cite{kim2017bistability} (10 and 100 nm in radius) obtained by the model. The comparison with experiments \cite{kim2017bistability} is reported. }
    \label{Raman_exp}
\end{figure}

In Fig. \ref{Raman_exp} the hLO mode for the two wires as given by the present model is compared with the data from  Ref. \cite{kim2017bistability}. 
We assume both systems to be large enough that we can parameterize (\textit{ab-initio} at $T=$ 0 K)  Eq. \ref{omega_1d} using the bulk $\epsilon_m$ and BECs, and purely longitudinal and constant eigenvectors. 
This results in an upper limit in terms of frequencies for each $q_z$ and $t$.\cite{SI} 
Shrinking the size corresponds to a decrease in the branch steepness close to $\Gamma$ and to the observed blue shift.  
The agreement between experiments and the model is very good for the thicker base of the nanowire and also semiquantitative with respect to the measured blue shift for the tip. There are multiple reasons behind this, such as temperature effects and mechanical contributions (not accounted for in this parametrization), but especially the uncertainty on the phonon momenta probed experimentally. However, the most reasonable explanation is that base and tip signals are not fully decoupled. Thus, experiments probe a mixture of base and tip phonons and
fall in between theoretical predictions for the two nanowires' size. In this regards, further comparison with experiments on single-wires/tubes with constant radius will serve to validate the details of the proposed model. To this aim, we hope this Letter will motivate future work in this direction.

In conclusion, we have argued for a breakdown of the dielectric shift experienced by LO phonons in 1D systems, and shown its exact asymptotic behavior in terms of phonon momenta and material size.
This novel understanding and the accurate first-principles description provided by the OBCs implementation of this  work represent a transparent and insightful advance in the field. First, the model can be exploited by the experimental community to aid material characterization when 1D systems are considered. 
In addition, the proposed computational framework unlocks the full potential of DFPT for 1D system, paving the way to further studies which go beyond vibrational properties and may revolve around electron-phonon coupling, and optical properties.

This work has been inspired by the experimental results from Ref. \cite{kim2017bistability}.
We aknowledge fundings from the Swiss National Science Foundation (SNSF -- project number 200021-143636) through  the  MARVEL  NCCR and the computational  support  from the  Swiss  National  Supercomputing  Centre  CSCS under project ID mr24. Fruitful discussions with  Anna Fontcuberta i Morral and Francesco Libbi are also gratefully acknowledged.

\bibliography{main}

\end{document}


\title{ Supporting information for: \lq Infrared-active phonons in one-dimensional materials and their spectroscopic signatures\rq}
\author{Norma Rivano}
\affiliation{\THEOSMARVEL}

\author{Nicola Marzari}
\affiliation{\THEOSMARVEL}
\affiliation{\PSI}

\author{Thibault Sohier}
\affiliation{\Montpellier}
\maketitle


\section{Analytical model}
In this section, we detail the analytical derivation whose results are summarized in the main text.
We model the 1D material by assuming a periodic charge distribution along the $z$-axis and an homogeneous charge distribution in the radial direction, confined within an effective radius $t$.
We assume vacuum outside, but the generalization to other surrounding media is straightforward.
$$
\epsilon^m=
\begin{pmatrix}
\epsilon_{p}&0&0\\
0&\epsilon_{p}&0\\
0&0&\epsilon_{z}\\
\end{pmatrix}
{Z}^a=\begin{pmatrix}
Z^a_{p}&0&0\\
0&Z^a_{p}&0\\
0&0&Z^a_{z}\\
\end{pmatrix}\,.
$$

\begin{figure}[h!]
    \centering
    \includegraphics[scale=0.35]{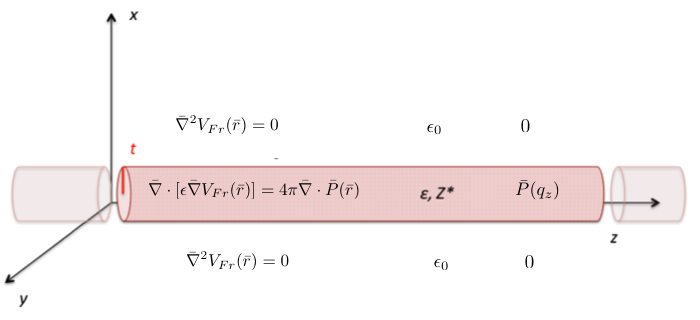}
    \captionsetup{width=\textwidth}
    \caption{ Schematic representation of a 1D polar material within our model, i.e., filled cylinder of radius $t$. The dielectric properties ($\epsilon^{1D}$, $Z^a$) are summarized both inside and outside the cylinder. The Poisson equations to be solved in each region are reported.}
    \label{sketch_model}
\end{figure}

Exploiting the periodicity along the $z$-axis, the LO atomic displacement pattern can be decomposed as a superposition of phonon waves of momenta $q_z$:
\begin{equation}
\bar{u}_{\mathrm{LO}}^a (\mathbf{r_p}, z)= \sum_{q_z} \mathbf{u}_{\mathrm{LO}}^a (q_z) e^{i {zq_z}}\,,
\label{dispalcement}
\end{equation}
where $\mathbf{u}^a_{LO}=  \mathbf{e}^a_{LO}/\sqrt{M_a}$ is the displacement of atom $a$ given by the eigenvectors of the dynamical matrix $\{{e}^a_{LO}(q_z)\}$ scaled by the nuclear masses $M_a$.
This generates the following Fourier-transformed polarization density:
\begin{equation}
\mathbf{P(q_z)}=\frac{e^2}{L}\sum_{a}{Z_a}\cdot \mathbf{u}^a_{LO} (q_z) f(\mathbf{r}_p)\,,
\label{polarization}
\end{equation}
where $e$ is the unit charge, $L $ is the unit-cell length in the periodic direction, $Z_a$ is the tensor of BECs associated to each atom $a$, and we model the out-of-chain profile as a Heavy-side step function in the radial coordinates, $\mathbf{r}_p$, $f(\mathbf{r_p}) = \frac{\theta(t-|\mathbf{r_p}|)}{\pi t^2}$ normalized to unity over the cross-sectional area  ($\int_{0}^{\infty} f(r_p)2\pi r_p, dr_p$=1).
Note that the same effective radius $t$ is used to define the extent of the dielectric medium (ions and electrons) and the polarization density.
The induced Fr\"{o}hlich potential, ${V}_\mathrm{Fr}$ (i.e., the same one related to the well-known electron-phonon coupling) is obtained by solving the Poisson equation:
\begin{equation}
\mathbf{\nabla} \cdot[\epsilon^m\mathbf\nabla{V}_{\mathrm{Fr}}(\mathbf{r})]=-4\pi\mathbf{\nabla} \cdot\mathbf{P}(\mathbf{r}) \,.
\label{poisson}
\end{equation}
The solution needs to fulfill the proper boundary conditions (i.e. continuity of the potential and its derivatives at the interfaces, and no divergences in $V_{\mathrm{Fr}}$).
The right-hand side of Eq. \ref{poisson} Fourier-transforms as $q_z\cdot P(q_z)$.
After imposing the continuity conditions, we find the potential ${V}_{\mathrm{Fr}}$ and we obtain the Coulomb screened interaction as
\begin{equation}
    W_c(q_z)=\frac{V_{\mathrm{Fr}}(q_z)}{q_z  P_z(q_z)}\,.
\end{equation}
We then derive the corresponding electric field $\mathbf{E}$ and the forces $\mathbf{F}_a$ acting on each ion $a$, being respectively:
\begin{equation}
    \mathbf{E}(\mathbf{q}_z)=-\nabla V_{\mathrm{Fr}}(\mathbf{q}_z)=-W_c(\mathbf{q}_z)(\mathbf{q}_z\cdot \mathbf{P}(\mathbf{q}_z)\mathbf{q} \,,
\end{equation}
\begin{equation}
    \mathbf{F}_a=\mathbf{E}\cdot Z_a=-W_c(\mathbf{q}_z)\Bigl(\mathbf{q}_z\cdot \frac{e^2}{L}\sum_{a'}Z_{a'}\cdot \mathbf{u}^{a'}_\nu\Bigr)(\mathbf{q}_z \cdot Z_a)\,.
\end{equation}
These new forces eventually turn into an additional contribution to the dynamical matrix:
\begin{equation}
D_{a i ,b j}(q_z)=  - {\frac{e^2}{ L}} W_c (\mathbf q_z)  \frac{(\mathbf q_z \cdot Z_a)_i (\mathbf q_z \cdot Z_{b})_j }{ \sqrt{M_a M_{b}} }\,.
\label{dynamical-matrix}
\end{equation}
where $a$ and $b$ label the atoms, while $i$ and $j$ index the cartesian coordinates.
This additional contribution is reflected in the eigen-solution of the dynamical matrix.
With an isotropic assumption on the already diagonal dielectric tensor ($\epsilon^m\to\epsilon_\mathrm{1D}\mathbb{I}=\epsilon_z\mathbb{I}$), this leads to the following phonon frequency for the LO mode:
\begin{equation}\begin{split}
    \omega_{\mathrm{LO}}=\sqrt{\omega^2_{0}+ W_c (\mathbf q_z)\frac{e^2|q_z|^2}{L}\biggl(\sum_a\frac{ Z_a\cdot\mathbf e^a_\mathrm{{LO}} }{ \sqrt{M_a }}\biggr)^2}
    =\sqrt{\omega^2_{0}+\Delta\omega^2_{\mathrm{max}} \Bigl[1- \Delta_{\mathrm{1D}}(q_z,t)\Bigr]}
    \label{omega_1d}
    \end{split}
\end{equation}

where the 3D prefactor $\Delta\omega_{\mathrm{max}}$  and the dimensionality modulation $\Delta_{\mathrm{1D}}(q_z,t)$ specific of 1D are reported in the main text:
\begin{equation}
\Delta \omega_{\mathrm{max}}^2= \frac{ 4 \pi e^2}{\epsilon_{\mathrm{1D}} \Omega} (\sum_a\frac{ Z_a\cdot\mathbf e^a_{\mathrm{LO}} }{ \sqrt{M_a }})^2 \,,
\end{equation}
\begin{equation}
\Delta_{\mathrm{1D}}(q_z,t)=2I_1 (|q_z| t) K_1 (|q_z| t) \Bigl(1 - \frac{2 \epsilon_\mathrm{1D} \sqrt\pi q_z t I_1(|q_z| t) K_0(|q_z| t) -G^{2 2}_{2 4}(|q_z|^2 t^2)}{2\sqrt\pi q_z t(\epsilon_\mathrm{1D} I_1(|q_z| t)K_0(|q_z| t)+I_0(|q_z| t)K_1(|q_z| t))}\Bigr)\,.
\end{equation}

Here we mention the limit behaviors. By by Taylor expanding  $\Delta_{\mathrm{1D}}(q_z,t)$ in Eq \ref{omega_1d} in the vicinity of $\Gamma$, we can extract the mentioned logarithmic overbending $q_z^2 log(q_z)$ ad the 3D limit (i.e., $q_zt \to  \infty$):
\begin{equation}
    \Delta_{\mathrm{1D}}(q_z,t) =\begin{cases}
        1-\frac{q_z^2t^2}{2}(C(\epsilon_\mathrm{1D})- \epsilon_\mathrm{1D}log(q_zt)) \, & \text{for } q_zt \to 0\\
        0 \, & \text{for } q_zt \to \infty

    \end{cases}
\end{equation}
where $C(\epsilon_\mathrm{1D})$ is constant with respect to $q_zt$  but it depends on the dielectric tensor and is equal to $-\frac{3}{4}+\frac{\gamma}{2}(1-2\epsilon_\mathrm{1D})+\frac{\psi(3/2)}{2}+log(2)(\epsilon_\mathrm{1D}+1)$.
The transition between the two limit behaviors is somehow more complicated with respect to what shown in 2D \cite{sohier2017breakdown}, with $(q_zt)_\mathrm{critical} \propto \frac{const.}{\epsilon_\mathrm{1D}}$.
\begin{figure}
    \centering
    \includegraphics[scale=0.4]{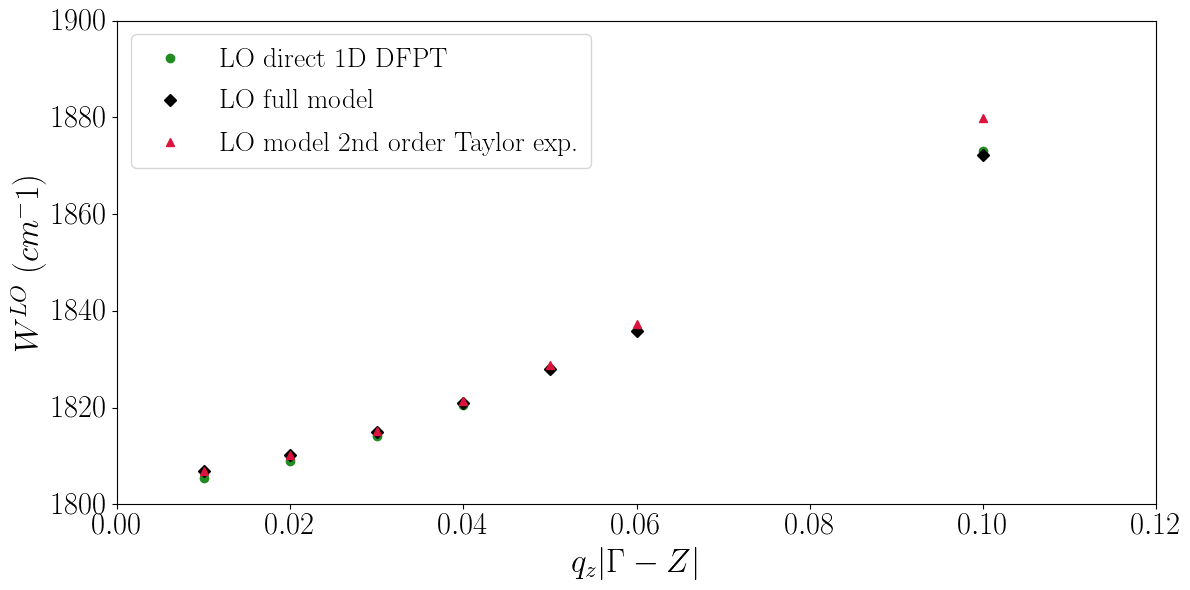}
    \caption{Long-wavelength limit for the LO branch of the BN atomic chain. Comparison between direct DFPT frequencies, the analytical model (full) and its Taylor expansion up to the second order in $q_zt$.}
    \label{fig:my_label}
\end{figure}

\section{1D open-boundary conditions in density functional perturbation theory}
Phonon calculations reported in the main text are performed with density functional perturbation theory (DFPT) in a newly implemented one-dimensional flavour. 
Using standard DFPT in 3D periodic boundary conditions (PBCs) issues arise for the LO dispersion at small but finite momenta due to spurious long-range interactions between the periodic images.
On the contrary, the computation exactly at $\Gamma$ happens to be correct in 1D, as in 2D, (zero dielectric shift) since the corresponding $\mathbf{q}=0$ macroscopic field is ill-defined and absent from the simulation.
The standard method of simply increasing the vacuum between images only reduces the affected portion of the BZ, while the computational cost significantly increases.
Thus, we propose to properly address this issue via the Coulomb cutoff technique \cite{ismail2006truncation,rozzi2006exact,castro2009exact, sohier2015density}. Our version is based on the one proposed in Ref.  \cite{rozzi2006exact}, is implemented in the relevant packages (PWScf and PHONONS) of the QE distribution \cite{RevModPhys.73.515, giannozzi2009quantum,giannozzi2017advanced} and soon to be released \cite{PRB}. 
This implementation leads to the correct 1D boundary conditions for the computation of total energies, forces, stress tensors, phonons, and the electron-phonon interaction. \

\begin{figure}[h!]
    \centering
        \includegraphics[scale=0.38]{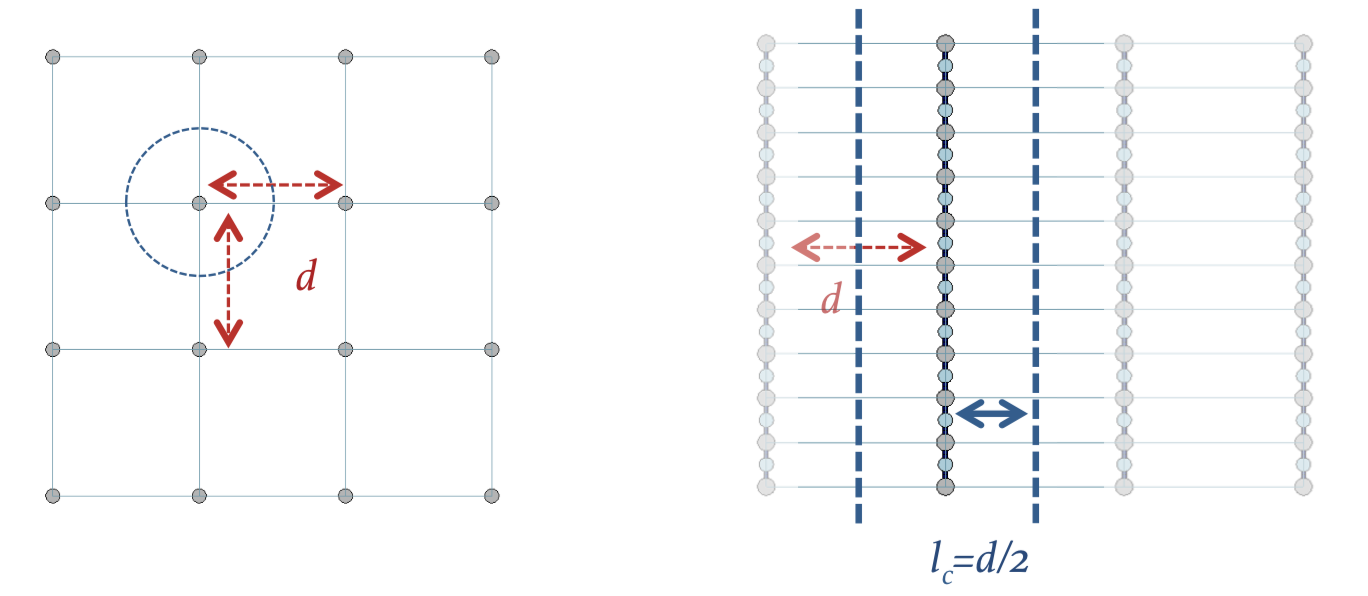}
    \captionsetup{width=\textwidth}
    \caption{Sketch of the supercell construction for 1D systems and the effect of introducing the Coulomb cutoff.}
    \label{sketch_model_pbc}
\end{figure}

A further in-depth publication will follow detailing the implementation and fully documenting its relevance in terms of long-ranged physics in 1D systems \cite{PRB}. 
Here, we report in a nutshell the main idea. The core strategy consists in redefining the Coulomb kernel $v_c \rightarrow v_c^*$ as:
\begin{equation}
    v_c^*(\mathbf r)=\frac{\theta(l_c-|\mathbf r_p|)}{|\mathbf{r}|}\,.
\end{equation}
All the relevant potentials are then obtained as convolution of this kernel with the charge density
   \begin{equation}
     V(\mathbf{r})=e\int \rho(\mathbf{r}') v_c^*(|\mathbf{r}-\mathbf{r}'|)\,d\mathbf{r}'\,,
 \end{equation}
in such a way that a given charge in the 1D system interacts only with charges within a cylinder of radius $l_c$ around it. Note that although the kernel is discontinuous, potentials are smooth thanks to the convolution with the charge density.
Eventually, the material is effectively isolated: we are left with a 1D periodic system, repeated in the two additional dimensions of space in order to build potentials that mathematically fulfill 3D PBC.
Note that the cutoff $l_c$ needs to be at least as large as the maximum distance between electrons belonging to the system, i.e. the effective thickness of the material $2t$, otherwise some physical interactions of the 1D system itself are erroneously cut. In turn, the size of the simulation cell in the non-periodic directions $d$ should be such that electrons belonging to different periodic images are separated by at least $l_c$.
In practice, the cutoff is chosen to be $l_c=d/2$, and the supercell  built such that $d>4t$.

\section{Computational details and model parametrization}
\subsection{Computational details}
The analytical and first-principles developments of this work are applied to the following systems: BN atomic-chain, armchair BN nanotubes of increasing radius (i.e., (4,4,),(5,5),(6,6)), and a small wurtizte GaAs nanowire passivated with H atoms to saturate the dangling bonds on the surface (24 atoms in total). 

First-principles calculations of structural properties and phonons are performed with the Quantum ESPRESSO (QE) package, by combining density-functional theory (DFT) and density-functional perturbation theory (DFPT) using the PBE exchange-correlation functional for all materials, with the exception of bulk wurtzite-GaAs, for which we use  norm-conserving pseudopotentials within the local density approximation from the 
Original QE PP Library. For 1D-DFT/DFPT calculations, the newly implemented 1D periodic-boundary conditions (1D cutoff and 1D phonon Fourier-interpolation) are applied to properly describe linear response to a phonon perturbation. While for 3D-DFT/DFPT (standard PBCs) we used the available QE distribution without further modifications.
Most of the pseudopotentials are taken from the Standard Solid-State Pseudopotentials (SSSP) library (precision version 1.1) \cite{prandini2018precision} and the wave-function and  charge density energy cutoff have been selected accordingly, being respectively: 110 and 440 Ry for the chain, 80 and 440 Ry for nanotubes, and 90 and 720 Ry for the GaAs nanowire. For bulk GaAs, we selected instead a wave-function cutoff of 80 Ry. We treated all the materials under study as non-magnetic insulator (i.e., fixed occupations) and a fine electron-momenta distance of approximately 0.2 $\mathrm{\AA}^{-1}$ (unshifted mesh) has been used to sample the BZ. 
The convergence of all the relevant parameters have been performed aiming to an accuracy on the final phonon frequencies of few $\mathrm{cm}^{-1}$.

\begin{figure}[t!]
    \begin{subfigure}{0.49\textwidth}
     \includegraphics[scale=0.4]{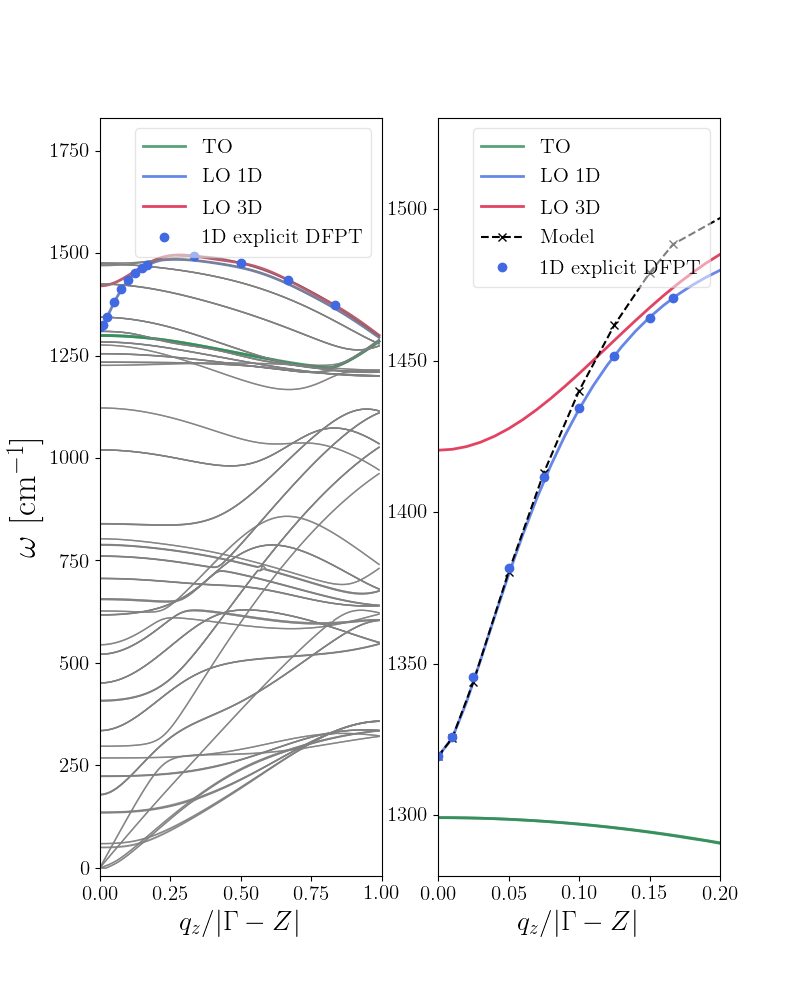}
    \end{subfigure}
    \begin{subfigure}{0.49\textwidth}
     \includegraphics[scale=0.4]{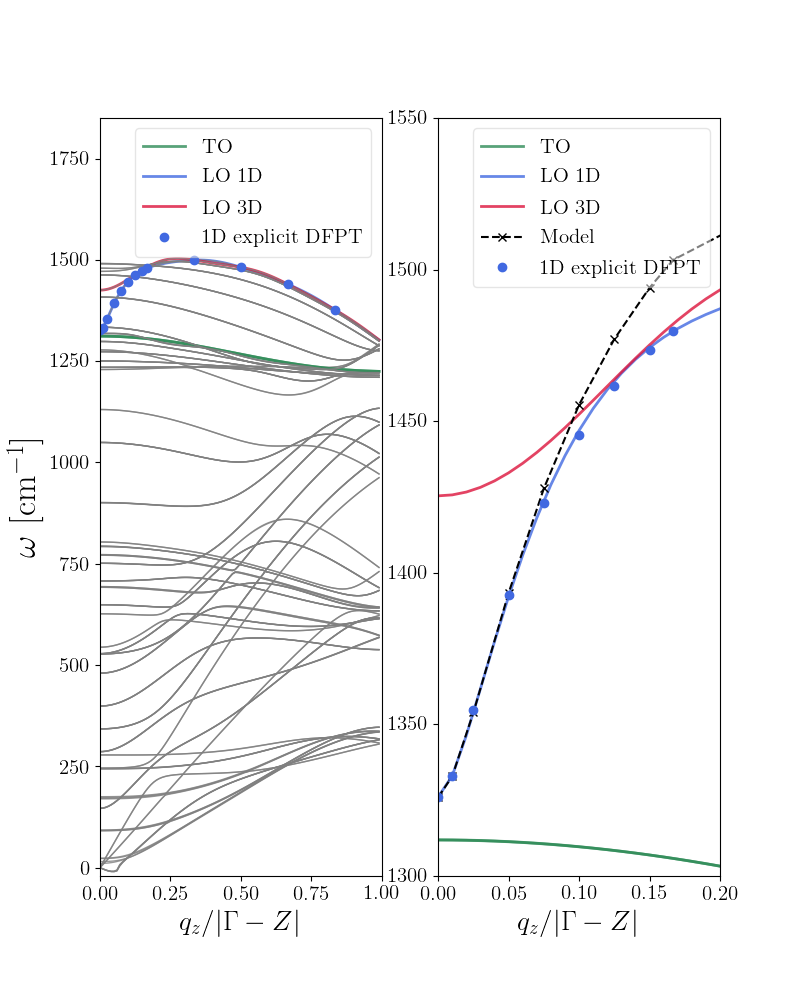}
    \end{subfigure}
    \caption{Phonon dispersion of (5,5) and (6,6) BN nanotubes, respectively. On the left panels, we compare the ab initio results using DFPT in 3D (no cutoff between period copies and 3D long-ranged electrostatic dipole-dipole interactions (LR) and using 1D DFPT (1D Coulomb cutoff and 1D LR). On the right, the interpolated and explicit DFPT long-wavelength polar optical phonons are isolated and compared with the analytical model.}
    \label{BN_nanotube_55}
\end{figure}

Here, we report fully in Fig. \ref{BN_nanotube_55} the phonon dispersion for BN-nanotubes (5,5) and (6,6), which are used for comparison to the (4,4) tube in Fig. 2 of the main text. Also, in Tab. \ref{tab} we summarized the central parameter of the model (i.e., the radius) and the LO and TO frequencies at $\Gamma$ to support the discussion on mechanical effects made in the Letter.

Referring to Fig. 1 from the main text, it is worth mentioning that the dispersion of the acoustic phonons in the proximity of $\Gamma$ are slightly negative and exhibit a characteristic \lq wing\rq.
These wings stem from violation of the invariance and equilibrium conditions on the lattice potential.
In fact, DFPT derived IFCS satisfy the proper conditions (depending on the dimensionality of the systems) only approximately due to numerical convergence issues such as insufficient k-sampling or incomplete basis sets.
The customary approach is then to enforce them, as a post-processing, via a specific set of relations called acoustic sum rules (ASRs) \cite{born1955dynamical, begbie1947thermal, PhysRevB.1.910}.
During the elaboration of this work, the suitable ASR, restoring the correct  quadratic dispersion for flexural phonons in 1D as well, has been implemented as described in Ref. \cite{changpeng}.

\begin{table}[ht]
\centering
\resizebox{0.6\columnwidth}{!}{%
\begin{tabular}{ c |  c | c | c | c | c}
\hline
\hline
1D-system    & t (bohr)       &   $\omega_\mathrm{LO}$ ($\mathrm{cm^{-1}}$)            &      $\omega_\mathrm{TO}$  ($\mathrm{cm^{-1}}$) & $\Delta \omega_\mathrm{mech}$ ($\mathrm{cm^{-1}}$) \\
\hline

BN-chain        & 1.70 &  1804 & 489 & 1315 \\

BN-NNT (4,4)    & 10.15 & 1306  & 1275 & 31 \\

BN-NNT (5,5)    & 11.49 & 1319 & 1299 & 20 \\

BN-NNT (6,6)    & 12.84  & 1326  & 1312 & 14 \\

GaAs-NW 24      &  10.41 & 244 & 240 & 4  \\

\hline
\end{tabular}%
}
\caption{Ab initio data and parameters related to the dielectric shift in polar 1D materials. $\omega_\mathrm{LO}$ and $\omega_\mathrm{TO}$ refer the values at zone center. $\Delta \omega_\mathrm{mech}$ designates the mechanical splitting at zone center. The effective radius $t$ is computed by averaging the electronic density profile in the out-of-wire directions and performing a gaussian fitting for the chain ($ t$=2$\sigma$), while setting a meaningful threshold of $\approx 10^{-5} $ a.u. for the other systems. }
\label{tab}
\end{table}

\subsection{Model parametrization}

The analytic results rely on the \textit{ab-initio} parameters obtained independently via DFPT in 1D open boundary conditions.
The implementation of Eq \ref{dynamical-matrix} requires several physical quantities.
Masses, eigenvectors, eigenvalues and Born effective charges are directly obtained from the underlying DFT and DFPT calculations.The only exceptions are the effective radius $t$ and the dielectric tensor $\epsilon^{\mathrm{1D}}$ which is trivially replaced by the dielectric component $\epsilon_z$ consistently with the isotropic assumption in our model, proved to be effective for our scopes.
As discussed in the literature \cite{sohier2016two, electric, dielectric, Libbi_2020, PhysRevLett.96.166801, tian2019electronic, andersen2015dielectric, mohn2018dielectric, sohier2021remote, sponza2020proper}, the dielectric tensor concept is ill-defined in nanostructures.
More advanced strategies have been proposed to model the response of two-dimensional materials \cite{andersen2015dielectric, mohn2018dielectric, sohier2021remote, sponza2020proper} based instead on the polarizability.
A particularly fundamental and robust theory has recently been proposed in 2D \cite{royo2020long}.
In this work, however, modelling the material as a dielectric cylinder implies the presence of a 1D dielectric tensor in our  model.
It differs from the one computed in QE $\epsilon^{\mathrm{QE}}$, which strongly depends on the size of the simulation cell. They are related via effective medium theory \cite{freysoldt2008screening, sohier2016two} as follows.
It is clear that the relevant physical quantity is the polarizability (longitudinal and transveral) which characterizes the response of the isolated 1D system per unit of length.
This quantity should be preserved in an isolated 1D model and in calculation with periodic images:
\begin{equation}
    A_{\mathrm{QE}}(\epsilon_{\mathrm{QE}}-1)=A_{\mathrm{1D}}(\epsilon_{\mathrm{1D}}-1)\,,
\end{equation}
where $A_{\mathrm{QE}}$ and $A_{\mathrm{1D}}$ stands for the cross-sectional area in the unit cell used by QE and in the 1D cylinder model, respectively.
In this case, the cutoff is expected to take care of the depolarization fields coming from the periodic images\cite{PhysRevLett.96.166801}.
The well defined physical quantity is the product $A_{1D}\epsilon_{1D}$.
Taken individually, $A_\mathrm{1D}$ and $\epsilon_\mathrm{1D}$ are just parameters of the model.
There is thus a level of arbitrariness in choosing $A_\mathrm{1D}$ and $\epsilon_\mathrm{1D}$ individually, keeping the product constant.
Being mostly interested the long wavelength limit $q_z \rightarrow 0$, we use the $A_\mathrm{1D} \to 0$ limit of this equation, while $A_\mathrm{1D}\epsilon_\mathrm{1D}$ is constant and determined by first-principles calculations.
We then get:
\begin{equation}
    \epsilon_\mathrm{1D}=\frac{c^2}{\pi t^2}(\epsilon_\mathrm{QE}-1)\,,
\end{equation}
where $c$ is the out-of-chain length characterizing the supercell-geometry (assumed to be the same in the x and y directions), and $t$ is the effective radius.
Within our model, $t$ characterizes both the electronic charge density and the out-of-chain polarization density profile.
Thus, a reasonable choice to determine this parameter would be to rely on the radial electronic charge density profile, averaged on the cross-sectional planes along the 1D-axis, and set a meaningful threshold.
In practice, in implementing the correction to the dynamical matrix, we automatize the choice of the radius as $t=d/4$, where in the most general case $d$ is the size of the cell in the non periodic direction.

An improvement of the agreement with DFPT could probably be  obtained, in general, by replacing the isotropic approximation with the full dielectric tensor (i.e., $\epsilon_z \neq \epsilon_p$). Then, for nanotubes in particular by modifying the choice made for the shape of the radial polarization density profile $f(\mathbf{r_p})$ in Equation \ref{dispalcement}.
 In fact, we are here proposing a filled-cylinder model, while nanotubes are more likely hollow-cylinders.
The reason why, despite this approximation, the model works  so well may be linked to the $p_z$ orbitals pointing inwards and thus filling the cavity of the tube.
The smaller the nanotube, the more it should resemble, in this sense, a filled cylinder for our purposes.
Nonetheless, since we are interested in the long-wavelength description of polar optical phonons, the agreement we get in the first tenth of the BZ is more than satisfactory even for the larger materials, and a further improvement is beyond the proposed scopes.

\section{Experimental relevance}

\begin{figure}
    \centering
    \includegraphics[scale=0.4]{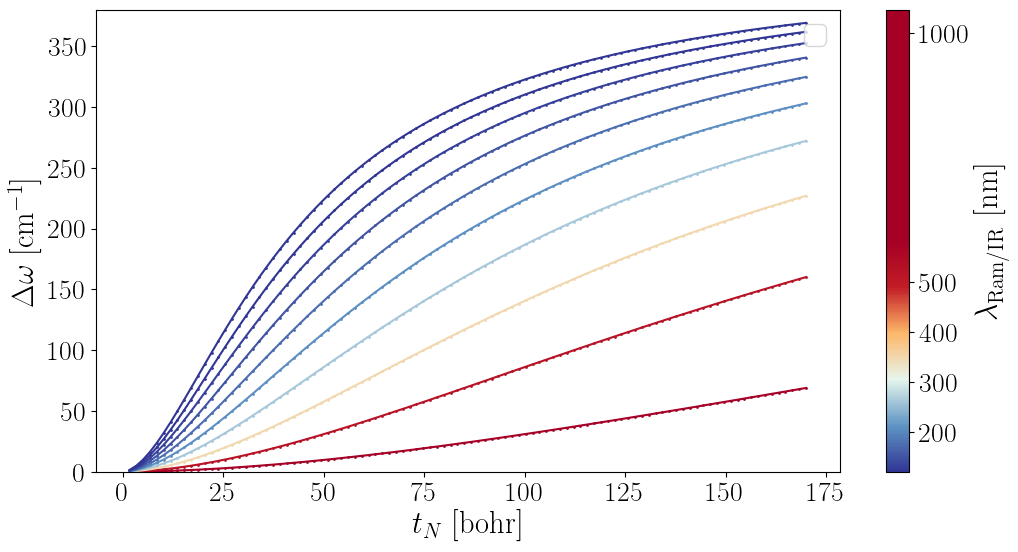}
    \caption{Evolution of the LO branch as a function of the radius of the BN chain (toy model) obtained by the model. The color bar indicates different Raman/IR wavelengths in experiments.}
    \label{raman}
\end{figure}

In the main text we have explained how the analytical model can be turned into a tool to complement/interpret photon scattering experiments on a variety of 1D systems. 
Here, it is instructive to follow in Fig. \ref{raman}  the hLO evolution with size for a toy-model: the BN atomic chain, modified by addition of atomic shells around its axis. The role of $t$ is symmetric with respect to the one of $q_z$, meaning that the presented plot is indeed similar to the standard phonon dispersion with the difference that instead of the phonon momentum $q_z$ across the BZ, on the x-axis we have a possible range of radii. Then, the color-bar corresponds to the experimental laser wavelength $\lambda$ which is related to the probed phonons by momentum conservation law (i.e., $q_z=2\pi/\lambda$) . The larger the $\lambda$, the closer to $\Gamma$ is the momentum probed. By increasing $t$, the logarithmic overbending gradually stiffens and tends to align parallel to the energy-axis, while the short-wavelength part of the \lq branch\rq becomes progressively flatter and wider, converging to the 3D limit given by $\Delta\omega_\mathrm{3D}$. These considerations are fully in agreement with what found for nanotubes in terms of DFPT. Similar considerations and possible usage of the proposed model hold as well for Raman/IR on single, isolated, and semiconducting wires/tubes.  

As a general comment, the model has been shown to perform remarkably well for different 1D prototypes (chains/tubes/wires) and different sizes. In this sense, the only caveat if one wants to use it as a complementary tool for Raman/IR experiments is the accuracy of the parametrization: radius, BECs, dielectric tensor, eigenvectors and reference frequency $\omega_0$.
Depending on the size of the system, two possible strategies are viable.
When dealing with small radii (like the BN-chain above), whose dielectric and mechanical properties are expected to differ from the ones of the bulk parent, the parametrization is a bit more delicate and computationally expansive. In fact, DFPT gives access to the above parameters at the cost of a single $\Gamma$ calculation. However,  the eigenvectors loose more or less rapidly their LO character  depending on the size, thus being strongly varying with both $q_z$ and $t$, thus requiring additional DFPT calculations for all the needed q-points involved in the spectrum 
On the other hand, for larger systems, e.g., the nanowires from Ref. \cite{kim2017bistability} a \lq top-\rq  approach can be appropriate. This basically consists in assuming that the basic ingredients of the model are the same from the bulk 3D or 2D parent. In this case, as commented in the main text, the mechanical size-effects are considered negligible and the dielectric ones are assumed to be fully conveyed by the radius explicitly appearing in the long-wavelength modulation $ \Delta_{\mathrm{1D}}(q_z,t)$.

As far as the comparison with Ref. \cite{kim2017bistability} is concerned, we further comment on the discrepancies between our predictions and experiments. 
Our model gives us a good agreement for the data from the base. Temperature effects, as well the mixture of wurtzite and zincblend phases in the experiment, could  help explaining the few $\mathrm{cm^{-1}}$ in excess. In fact, the LO (TO) Raman peak at 12 K is reported to be $293\pm 1 \,\mathrm{cm^{-1}}$ ($271\pm 1 \,\mathrm{cm^{-1}}$) \cite{strauch1990phonon} which is in agreement with our predictions at 0 K), while at 296 K is $285\pm 7 \,\mathrm{cm^{-1}}$ ($267\pm 3 \,\mathrm{cm^{-1}}$)\cite{waugh1963crystal}, closer to Ref. \cite{kim2017bistability}.
Also, the branch is extremely steep in this case and the steepness seems to match well with the experimental value reported, i.e., the experimental data lies on the line of the model even if for slightly different $q_z$ (see Fig. 3 in the Letter).
Focusing on the smaller nanowire (i.e., the tip), the difficulty to focus exactly the laser spot seems to be the most evident explanation behind the discrepancy: the signal from the tip and the base are not fully decouples and we end up probing a mixture of signal between the two sizes. In this sense, to assess the quality of our predictions for a wide range of radii (i.e., various steepness of the LO branch) we would need new experiments on single 1D systems, isolated an constant in size. In fact, a constant radius would avoid this dependency of the measured frequency on the focus of the laser. The isolation is instead fundamental to avoid cross-talks between the systems, which are instead present if the spot size of the laser is comparable to the inter-systems distance: in this case, for every phonon $q_z$ smaller than the inverse of this distance, the measure the response of a 3D weakly bounded systems made up by the different wires/tubes/chains. 
Besides these main requirements, also a good control on the laser wavelength and the general set-up is needed: the non perfectly mono-chromatic nature of the laser (dependent on the type of source used) affects the set of $q_z$ phonons effectively excited and thus contributing to the final spectrum. Similarly, depending on the penetration-depth of the exciting radiation, low-dimensional materials are known to experience a relaxation of the Raman selection rules, thus contributing to additional broadening of the spectrum. Temperature and laser heating are also crucial parameters affecting the final spectra. 
In conclusion, to fully validate the model and use it as simple tool for the experimental community, a stronger interaction between the two communities is desirable. 

\bibliography{main}